\documentclass{article}

\usepackage{booktabs}
\usepackage{url,graphicx,xxx}
\usepackage{enumerate}

\usepackage{setspace}
\usepackage{xspace}
\newcommand{\ie}{\emph{i.e.}\xspace}
\newcommand{\eg}{\emph{e.g.}\xspace}
\newcommand{\etc}{\emph{etc.}\xspace}

\begin{document}
\title{Challenges in the capture and dissemination of measurements from high-speed networks}

\author{
    R.~G.~Clegg$^{1,2}$ 
         \and
    M.~S.~Withall$^3$  \and 
    A.~W.~Moore$^4$  \and 
    I.~W.~Phillips$^5$  \and 
    D.~J.~Parish$^3$ \and 
    M.~Rio$^2$ \and
    R.~Landa$^2$\and 
    H.~Haddadi$^2$  \and 
    K.~Kyriakopoulos$^3$\and 
    J.~Aug\'{e}$^4$\and 
    D.~Salmon$^6$ 
}

\maketitle
\begin{abstract}
The production of a large-scale monitoring system for a high-speed network
leads to a number of challenges. These challenges are not purely techinical
but also socio-political and legal. The number of stakeholders in a such a
monitoring activity is large including the network operators, the users, the
equipment manufacturers and of course the monitoring researchers.
The MASTS project (Measurement at All Scales in Time and Space) was created to
instrument the high-speed JANET Lightpath network, and has been extended to 
incorporate other paths supported by JANET(UK). 

Challenges the project has faced have included: simple access to the network;
legal issues involved in the storage and dissemination of the captured
information, which may be personal; the volume of data captured and the rate
at which this data appears at store.
To this end the MASTS system will have established four monitoring points each
capturing packets on a high speed link. Traffic header data will be continuously
collected, anonymised, indexed, stored and made available to
the research community.  A legal framework for the capture 
and storage of network measurement data has 
been developed which allows the anonymised IP traces to be used for research purposes.

\end{abstract}

\footnotetext[1]{Corresponding author: richard@richardclegg.org}
\footnotetext[2]{Electronic \& Electrical Engineering,
            University College London}
\footnotetext[3]{Electronic \& Electrical Engineering,
        Loughborough University}
\footnotetext[4]{Computer Laboratory,
        University of Cambridge}
\footnotetext[5]{Computer Science,
        Loughborough University}
\footnotetext[6]{JANET(UK)}

\setcounter{footnote}{6}

\section{Introduction}
\label{sec:intro} 

The common availability of quality monitoring hardware, high-performance
computers and a ready supply of interesting network-use has led
the research community to somewhat become blas\'{e} about network
monitoring. However, a short discussion with any practitioners of
network-monitoring reveals that the topic is both complex and fraught. We
intend this paper to serve two purposes.  Firstly it provides a roadmap,
a commentary and insight for future contributors in the monitoring field
and secondly it describes a data resource which will be of great use to
the network modelling and analysis community.

The MASTS project (Monitoring at All Scales in Time and
Space)\footnote{\url{http://www.mastsproject.org}} is an EPSRC funded
collaborative research project between three universities: Loughborough,
Cambridge and University College London (UCL).  The project aim was to create
and operate a monitoring system for various links JANET (the network which
carries traffic to and from the UK academic community) and JANET Lightpath
(which carries research and scientific data).  JANET Lightpath is a 10Gb/s
network, previously known as UKLIGHT, which is operated in the UK by
JANET(UK), previously known as UKERNA.   It supports a range of research
activities and carries traffic from a number of Grid research projects.
In the MASTS project the JANET Lightpath network provides both a system
to monitor and backhaul for the collected data.

MASTS aims to provide information to network operators, network users
and network researchers.  The solutions reached to the challenges of
monitoring high performance networks as addressed by MASTS offer a
systems level set of solutions to communication network monitoring and
solutions covering the monitoring interface, storage and legal aspects are
presented in this paper. The project has also investigated solutions to
the visualisation, compression and analysis of monitored network data, but
these aspects are reported elsewhere~\cite{withall2007,Kyriakopoulos2007}.

The ultimate aim is to provide to researchers a database of layer two,
three and four header information for four monitoring points on the
network, three of which are carrying scientific/technical data on the
JANET Lightpath network and one of which is carrying data on the main
JANET network.  An anonymised version of this data is made available
to all researchers who sign an Acceptable Usage Policy.  The data sets are
catalogued in a searchable database and enhanced with metadata.

The internet, once a mere research-vehicle, now forms the background for
substantial parts of the economy and is fundamental to much social
intercourse. Improving performance of broadband IP networks have been
central to this with IP networks able to carry any data type. The
heterogeneity of IP networks, their ability to carry a \emph{triple-play} of
services (Television, telephone and data-services) to every broadband
consumer has lead many ISPs to transition to IP-based national backbones
(\eg British Telecom's 21st CN\footnote{\url{http://www.btplc.com/21CN/}})
and will motivate movement to an IP-based network at the foundation of
all communications services. Our system, aimed at 10Gb/s, is ideal for
monitoring the current-generation backbones and next generation
distribution-networks of such new broadband networks.  Understanding drawn
from MASTS will permit both a better understanding and more
sophisticated optimisation of an IP-based world.

\subsection{Background and Motivation}
\label{sec:motivations} 

Many researchers approaching network monitoring with a need for network
data (perhaps to validate a theory or provide input to a simulation or
study) quickly find themselves overwhelmed by the complexities of
monitoring.  Performing meaningful monitoring operations on high
performance networks is a complex challenge, which embraces not only the
technical issues of connecting to a network and storage of the information
collected, but also the procedural and legal issues of allowing this
information to be disseminated to interested users world wide.

Network monitoring covers a vast spectrum of activities from using
\emph{tcpdump} or \emph{wireshark} on a personal computer and understanding
why your browser is misbehaving, through to the wide-area monitoring
of data flows across an entire ISP as input to auditing, accounting
or network intrusion detection. However, in all but the most trivial
network-monitoring, the researcher will need to interact with the
operators of the relevant network. When such networks are in-house this
can make the process easier.  However, there is no guarantee. Network
operations staff are focused on the day-to-day and longer-term operational
needs of a network; researchers wishing to monitor networks --- often
focused on their own research-deliverables --- may only distract from
the day-to-day operations and are commonly seen as a \emph{tax} upon
operators time and resources. Such diverging interests are not the only
trap in network-monitoring; there is a vast array of different legal
and technical challenges to be faced~\cite{paxsonsound}.  

A number of research groups and organisations have developed
network monitoring systems including CAIDA~\cite{claffy2000},
RIPE~\cite{zhou2005} and NLANR (which is no longer operating). Many of these projects
use active measurement whereby specific packets are generated and added
to the existing network traffic.  In such scenarios, a
degree of control exists with respect to the rate at which measurements
are made. Many of the probe designs such as those produced by the RIPE
Test Traffic Measurement project\footnote{\url{http://www.ripe.net/ttm/}}
are intended for use on multiple internet paths and provide results for
many different paths at a low rate. However, the lack of performance
monitoring and diagnostic mechanisms has been highlighted in several
places~\cite{nrc2001looking,gap}.

The MASTS project therefore has designed passive probes for use on
10Gb/s links\footnote{Throughout this work we will use B for bytes
and b for bits, 10Gb/s represents 10Gbits/s }. As such, node deployment
is sparse compared with many other projects but more data is generated
from each probe. It is generally recognised that network monitoring is a
complex, multidisciplinary activity requiring the optimisation of many
parameters. Some of these issues have been addressed by CAIDA~\cite{murry2001}; 
whereas this paper presents the main issues and
solutions as seen by the MASTS project.

From the outset, the project took to heart the adage \emph{good
data outlives bad theory}~\cite{keeling1998rewards,nrc2001looking}.
A long-term archive of activity in the JANET Lightpath network was
planned.  The monitoring system was designed to cope with a growing
network and the database system is intended to provide a long-lived
resource to the community.

Many worldwide projects exist that collected and/or disseminated
packet-level traces, \eg the previously mentioned CAIDA and
NLANR \footnote{\url{http://pma.nlanr.net/}} projects, the CRAWDAD repository
\footnote{\url{http://crawdad.cs.dartmouth.edu/}} and
the Bellcore project \cite{leland93selfsimilar} are all good examples.
The aim of MASTS is to complement these data sets with data from faster
links, provided online soon after it is generated.  The availability
of this data and extra-derived data (we keep both packet traces and
aggregated flow information for longer periods) is crucial for several
communities. Traffic analysis, long range dependency, fault analysis,
denial-of-service detection, \etc can all profit from a large data set
representative of a significant Autonomous System.

\subsection{Legal Issues}
\label{sec:legal}

The very process of passive network monitoring involves the capture
(and usually) storage and analysis of information generated by users
other than those involved in the monitoring process itself. Potentially,
this can lead to serious legal issues associated with privacy and data
protection. This
situation is generally influenced by some or all of the following
characteristics:
\begin{itemize}
\item the purpose of the monitoring operation;
\item the ownership of the data so collected and its location;
\item the anonymisation approach  adopted;
\item the nature of the data to be collected, including the protocol layers;
\item the sources of the data and
\item the form of the data to be stored and disseminated.
\end{itemize}
In order to manage the legal status of the monitoring activities,
the particular combination of these characteristics determines the
legal status of the monitoring activity and the liable parties for
any abuse.  The approaches developed by the MASTS project are discussed
in Section~\ref{sec:legalframework}.

\section{Architecture}
\label{sec:arc}

The architecture of any network monitor is largely informed by the link to
be monitored, the constraints of cost and the objectives the project may
seek to optimise. In the case of the MASTS project the intention from the
outset was to design capture systems that perform full line-rate capture.
This is not to imply capture every octet of every packet will always
be captured.  However, a system was desired that was engineered to allow
as close to this as technically and legally permissible.

The first hurdle was to design a monitoring system to the physical
interface of the network-link to be monitored.  The opportunity of the
JANET Lightpath project, a new network infrastructure, provided a unique
chance to build a monitoring system in concert with a specific physical
infrastructure. Network practitioners will recognise that there are
numerous ways a particular link/capacity may be provisioned. A range
of physical options (copper, fibre, wireless) along with a range of
data-link-layers (SDH,  packet over SONET, raw (LAN) Ethernet) and
a wide range of speed options led to a huge number of alternatives,
each with it's own cost and benefit.

By being involved with the operational-deployment from the outset
the project allowed for fibre (interception)-needs and space-needs
be accommodated, while keeping the monitoring team appraised of
the operational-network's deployment.  The physical links of the
infrastructure are capable of 10Gb/s, however, the majority of the
installation was based upon a specific vendor's proprietary SDH frame
format.  We could not monitor these links using a splitter alone. This
led to two different solutions: one for parts of the JANET Lightpath
network with an alternative approach for other monitoring installations.

The physical interconnections dictated two different approaches to
present data from the three different links being monitored:
\begin{itemize}
\item JANET Lightpath: dedicated line card;
\item JANET Lightpath RAL-CERN: 10Gb/s splitter and
\item JANET internet interconnect: 10Gb/s splitter.
\end{itemize}

Aside from physical interconnections the specific project goals led to
an architecture optimised to minimise uncontrolled loss while allowing
best control over the long-term archiving of data.

\subsection{Physical Architecture}
\label{sec:physarc}

\begin{figure}[ht!]
\begin{center}
\includegraphics[width=\textwidth]{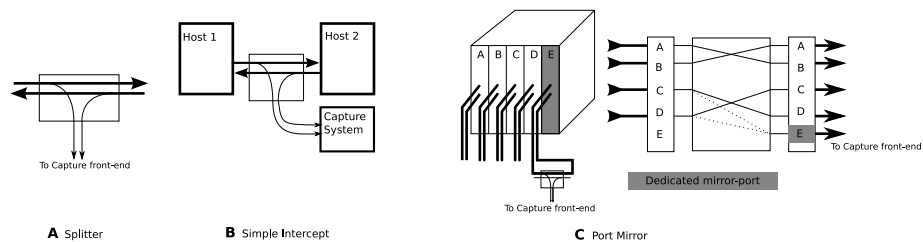}
\caption{Intercepting data for capture.}
\label{fig:splitter}
\end{center}
\end{figure}

Traffic interception is subject to constraints in both the political and
engineering fields. While the political considerations
are discussed elsewhere, we describe the two physical solutions
employed in our implementations. Clearly the design of any capture
system is tightly coupled to the physical media. In the case of
MASTS, some physical media installations did not lend themselves to
interception. To be economically intercepted\footnote{Without the
need to construct special purpose equipment running to millions of
pounds.}, the physical line representation needs to be able to be
interpreted by monitoring hardware (best thought-of as enhanced
interface boards.) This is entirely practical when the physical
line-encoding is one of a number of standards: for 10Gb/s the relevant
IEEE Ethernet agreed standards are:
\begin{itemize} 
\item the 10Gb/s LAN PHY: a physical layer for use in short-haul
networks and
\item the 10Gb/s WAN PHY: a physical layer for use in longer-haul networks
and compatible with common telecommunications (SONET/SDH) equipment.
\end{itemize}
Figure~\ref{fig:splitter}~\textbf{A} provides a diagrammatic
representation of a splitter internal: for each direction of flow a
percentage\footnote{While 50:50 splitters may be used, more usually
80:20 or 90:10 where the majority of the photons are not intercepted, is
common practice.} of light is redirected to a second output. Splitting
the light-flow in each direction provides two flows of data from the
intercepted physical interface.

Figure~\ref{fig:splitter}~\textbf{B} provides illustration of a
trivial intercept: collecting data flowing between \emph{host1}
and \emph{host2}. The splitter provides intercepted data for the
\emph{Capture System}, the hardware of the capture system may range from
simply a pair of unused network interface adapters through to dedicated
capture hardware. The differences between a simple solution and a more
sophisticated approach, such as that described here, relate to the
accuracy of time-stamping within the capture system.  Standard network
interface cards have not provided an accurate timestamp, sufficient
card capture facilities to minimize loss.  Buffer memory is a critical
resource to overcome bandwidth limitations in a computer architecture,
(often orders of magnitude more than that provisioned on a regular network
interface card).  Alongside this, the network interface card needs to
provide appropriate hardware and software support for the most efficient
mechanisms to move data into the capture system. Given our approach is,
at first approximation, to capture all data on the physical interface we
do not need the ability to selectively filter and discard irrelevant data
(a feature often present on network hardware).

While standards such as the LAN PHY 10Gb/s Ethernet and the WAN PHY
10Gb/s Ethernet are common, the physical presentation may not follow
such an open standard; such is the case for some of the links within
JANET Lightpath.  This led to a rather different solution for one
of the MASTS monitoring systems; a dedicated monitoring port in the
network infrastructure is used to mirror traffic from particular
ports. Illustrated in Figure~\ref{fig:splitter}~\textbf{C},
this approach may be recognizable to readers as similar to
the \emph{Switched Port Analyzer} (SPAN) on Cisco switch
equipment\footnote{\url{http://www.cisco.com/warp/public/473/41.html}}.

The project's use of port-mirroring differs in several
important ways.  Firstly, in a switching infrastructure the use of
port-mirroring may lead to high levels of packet jitter and packet
loss~\cite{zhang07traffic}. Secondly, it is important to over-provision
the monitoring port.  Clearly monitoring a 1Gb/s connection will
require 2Gb/s of monitoring capacity (1Gb/s for each direction). For
architectural reasons these two problems have limited impact on our
use of port-mirroring in JANET Lightpath. The port-mirroring activity
is done by a TDM (time-division multiplex) switch at the TDM level,
this means that the timing relationship between packets in a single
direction is undisturbed and the timing-error between packets of each
direction within a multiplex is a small bounded number of the order of
a TDM slot-length (\eg 15.625$\mu$s); thus it may be easily corrected in
the capture system. The second issue of over-provisioning is addressed by
this approach being limiting to the monitoring of at most 5~full-duplex
circuits.  Within the JANET Lightpath service each circuit is 
typically provisioned at 1Gb/s and most services are based upon 
these 1Gb/s circuits (although finer grained provisioning is possible).

In this particular configuration a port loopback is used and a splitter
is employed to extract the intercepted data-stream. 
This is because the intercept board does not provide any input data. The
capture board has no reason to transmit data and thus has no 10Gb/s
laser. However, in-common with much telecommunications equipment,
without a valid input the monitoring port will not initialise and send
any data. One solution is to use the loopback, sending the monitor-port
data back into the monitor-port. The switch will not actually process this
data as no paths are configured from the monitor-port to any destination.
This eliminates the risk of (unintentionally) injecting replica junk 
traffic back into the switch.

\begin{figure}[ht!]
\begin{center}
\includegraphics[width=.6\textwidth]{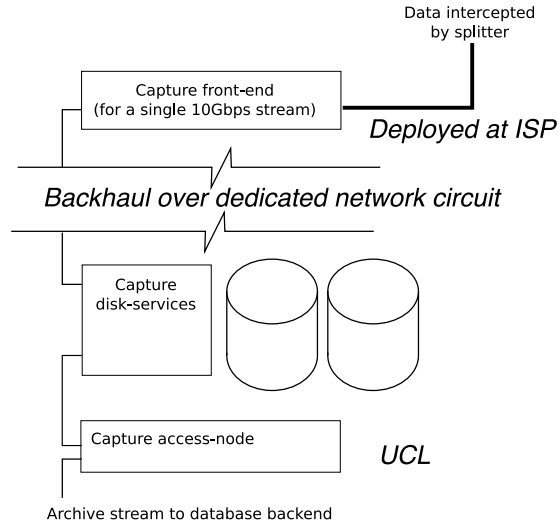}
\caption{The structure of the 10Gb/s monitor elements used within the
  MASTS project.}
\label{fig:collectarch}
\end{center}
\end{figure}

Once intercepted, packets need to be stored, processed and passed to
the database back-end without loss (or at least loss-limited)
continuously. Capture systems in the past have
often operated in a \emph{capture to local disk} for a period
and then off-line they would move or process relevant data. As
noted above, continuous capture was a driving imperative for this
architecture.  Figure~\ref{fig:collectarch} illustrates the capture
system we employ. The physical architecture is optimised toward a
lossless capture of all packets on a particular physical link.  This
means adequate provisioning of intermediate storage is needed throughout
the capture system. As any student of telecommunications systems will
know, data will require buffering at every point that throughput maybe
discontinuous. These discontinuities are the interfaces between parts
of the capture system as well as the parts of the capture system where
data-processing may, for short periods, exceed available resources.
Data is intercepted using an Endace DAG6.2SE Network Monitoring Interface
Card in a dedicated Dell PowerEdge 2850. While the capture card is
capable of receiving 10Gb/s, legal constraints restrict the capture
to only the transport/network headers; payload is removed. As noted
in Section~\ref{sec:dataavailable}, this significantly reduces the
required bandwidth.  Even in the worst case (a continuous stream of
the smallest packets) the throughput requirements are well within the
specification of the host machine. The host machine must move the data
(captured packets along with timestamps) from the capture card 
to intermediate storage. Along with the captured data,
the host machine also logs metadata related to the health of the capture
hardware, the host machine and so on.

In our architecture a SAN (System Area Network) is
employed that allows tight coupling of the capture-card
host. The SAN permits low-overhead/high-performance
sharing of manipulated files and is based upon the GFS cluster
filesystem\footnote{\url{http://sources.redhat.com/cluster/gfs/}}~\cite{soltis96gfs}
over the ATA-over-Ethernet interface~\cite{cashin05aoe}.  The storage
disks of the SAN provide access to the captured data (and associated
meta-data) through ancillary machines. The use of a SAN provides coarse
grained control of priorities which in-turn allows the capture system
writing new data to always have priority writing new data to the SAN over
any unduly heavy data-read operation.  The current (over)-specification
of hardware can accommodate the 10Gb/s stream, further, with the use
of the intermediate disk, the system has significant local storage
capacity allowing buffering of captured data if the down-stream nodes,
(capture access-node) require rebooting or have become CPU-bound in
tasks such as the anonymization of headers. Like any SAN, there is no
reason why multiple access-nodes can not read data from the SAN storage
if required; this may prove particularly useful if intermediate process
tasks such as the anonymization of headers (Section~\ref{sec:anon})
required multiple machines.

Captured data and log data formats consist of regular data-files
with a strict, pre-agreed naming convention incorporating the time of
capture. The capture system employs a fixed upper size, however, the
capture system also has a maximum period of time to wait before capture
and log files are rotated (closed, renamed and re-opened). In this way
a steady upper and lower-bounded stream of information can be guaranteed
to be made available from the capture system to the database back-end.

\subsection{Database Architecture}
\label{sec:dataarc}

\begin{figure}[ht!]
\begin{center}
\includegraphics[width=\textwidth]{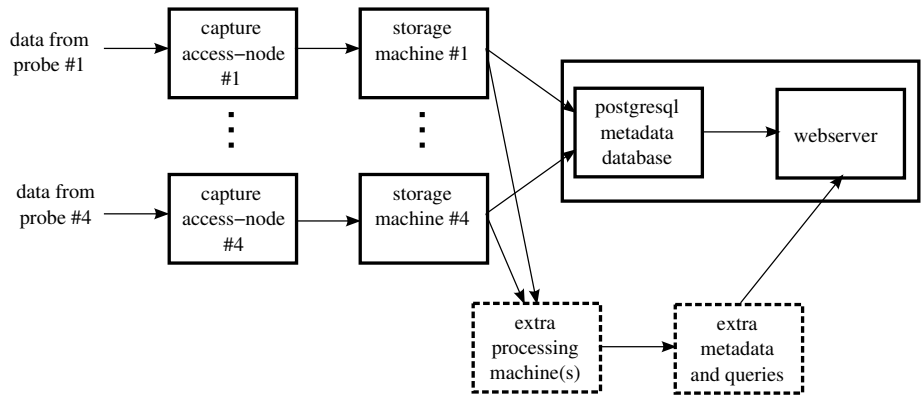}
\caption{The machines which are deployed to collect and analyse the
data.}
\label{fig:dbarc}
\end{center}
\end{figure}

An overview of the physical architecture is shown in
Figure~\ref{fig:dbarc}.  This shows the systems currently
in place at UCL and (using dotted lines) 
those planned additions later in the project.
In the current deployment the webserver
and database are on the same physical machine.

Once the capture system has finished writing a trace file (in ERF
format) and its associated metadata to the archive, the IP addresses
are anonymised (see Section~\ref{sec:anon}). The accompanying metadata
contains information including the time window covered by the trace file,
the monitoring point and several basic statistics such as number of
packets and bytes captured. In addition to the per trace file metadata,
probe configuration and monitoring point information is provided
out-of-band with the packet capture process, in the form of an XML
file. This includes information about the hardware and software used,
which link is being monitored and the bandwidth of the link.  Both the
trace file metadata and capture system information are inserted into
a PostgreSQL database\footnote{\url{http://www.postgresql.org/}}. This
database is suitably indexed to allow trace files to be found and simple
statistics to be derived. In conjunction with the database importing
system is an archive disk management system, which handles removal
of expired trace files (although some metadata is still retained for
removed files).

External users can search though and access the trace
file archive via a web-based interface (written in Python
TurboGears\footnote{\url{http://www.turbogears.org/}}).  Before accessing
the archive, the user must first register and accept the terms and
conditions of use (see Section~\ref{sec:legalframework}).  Only registered
users may download the trace files. Once the user has registered they
are issued a unique username and password for accessing the web-based
interface. Within the interface users can search for files by link,
probe, time or other combinations of the metadata. The resulting
trace files can then be downloaded as ERF files (tools are provided
to convert to the more standard pcap format). In
addition to searching for and downloading, metadata visualisations can
also be created (such as graphs of throughput for a particular link and
time period). Further preprocessing and visualisation capabilities are
planned (see Section~\ref{sec:conc}).

\subsubsection{Anonymisation}
\label{sec:anon}

A network-researcher may ideally wish to access a high-fidelity network
trace where the payloads of the data indicate clearly the activity
of the users and IP addresses easily identify end-points in the real
world.  However, implications of the legal constraints control the
data accessible: requiring payload data be removed and, in the United
Kingdom, the end-users not be identifiable.  These needs lead us to
our anonymization process. We do not capture payloads at the capture
system. The removal of payloads improves the performance of the capture
system; Section~\ref{sec:dataavailable} illustrates the significant
difference in the raw data-rate of captured data that discarding payloads
can provide.  When engineering a capture system it is thus advantageous
to discard payloads at the capture point reducing the quantity of data
to be managed within the capture architecture.

Aside from the removal of payloads, the industry standard
Crypto-Pan~\cite{fan2004prefix} is employed to provision a prefix-preserved,
anonymised IP address. Preserving address prefixes maintains the structure
of the IP address allowing for studies of routing and identifying
groups of end-systems but removes information permitting the specific
identification of a user, thereby satisfying the legal constraints.
Users are required to sign an acceptable use policy forbidding attempts
to reverse engineer the anonymisation before downloading the data
(see Section~\ref{sec:legalframework}).

\section{Practical Implications}
\label{sec:results}

The results of a project such as MASTS are varied and not limited to
purely measurements.  In reality, as the project has had to interact
with real operators on real networks results include documentation
of these interactions.  This section details the issues we have had
operationally and legally.  Some initial results follow.

\subsection{Operational Issues}  
\label{sec:operational}

As with any monitoring and measurement project a significant problem is
issues arising when working on real networks and with their operators. As
those responsible for the running of the network, operators need to
ensure that the user service is always supported. In this section the
common practical problems in network monitoring are examined and the
solutions for the MASTS project enunciated.

\begin{description}
\item[Availability] -- The primary purpose of a network is to provide a
connectivity service, and thus the primary purpose of the operator is to
ensure that the connectivity remains.  A common monitoring method is to
insert an optical splitter into the fibre to take a copy of the traffic.
Such an operation has two consequences: firstly that the fibre will
need to be broken, with subsequent loss of service; and secondly a fear
that the drop in signal will affect traffic. It is quickly apparent
that \emph{at-risk} maintenance periods need to be scheduled for such
installation and testing --- this requires a comfortable relationship
with the network-operations staff.

\item[Standards] --  Although a number of common standards exist
for interoperability between different suppliers, it is most common
for a network to be constructed from a single supplier's equipment.
This usually allows for the use of specific non-standard, proprietary
extensions and this caused problems in getting data from
parts of the JANET Lightpath network (see Section \ref{sec:physarc}).
To overcome this a novel hardware solution was necessary and the
difficulties of obtaining, installing and configuring 
cutting-edge monitoring hardware
(and consequent delays to the project) should never be under-estimated.

\item[Operator Cooperation] -- Placing a new card into an operational
switch requires a number of considerations.  When the researchers are
not operators of the network to be monitored the problem becomes far more
complex than simply purchasing monitoring hardware and plugging it into
a rack.  Often purchase, installation and configuration will need the
active co-operation of the network operator and this can lead to delays
at each stage.  One solution to be considered for future projects is
the \emph{embedding} of a project member within the network operator

\item[Data Storage Requirements] -- The project proposed 
monitoring several bi-directional 10Gb/s links.
Obviously this produces an enormous amount of data.  The issues
involved with storing this amount of data are discussed
in Section~\ref{sec:dataavailable}.

\end{description}

\subsection{Legal Framework}
\label{sec:legalframework}

In order to manage the legal status of the monitoring activities, the
MASTS project recognises 5 different categories of user or organisation
for its monitoring operations:
\begin{enumerate}
\item\label{en:class1} The Network Operator (JANET(UK) in this case).
\item\label{en:class2} The organisation holding the monitored data 
(UCL for this work).
\item\label{en:class3} Other MASTS project members using the data 
(researchers at Cambridge and Loughborough).
\item\label{en:class4} External users using packet level data.
\item\label{en:class5} External users using summary data.
\end{enumerate}

It was necessary to establish different agreements for each of the above
groups due to the differing legal nature of the relationships.  
Agreement ~\ref{en:agg1} is between ~\ref{en:class1}
and~\ref{en:class2} and covers their relationship.  Users
in~\ref{en:class3} are covered by ~\ref{en:agg1} as
well by the mechanism specific in ~\ref{en:agg2}; 
agreement~\ref{en:agg3} covers users in
category~\ref{en:class4};
finally, users in category~\ref{en:class5} are not covered
by an explicit legal agreement because they only have access
to summary data with is not 

\begin{enumerate}[A.]
\item\label{en:agg1} A legal agreement between the Network
Operator and UCL as the site holding the data. The resulting
document\footnote{\url{http://www.mastsproject.org/legal.html}}
establishes a practical example of a monitoring agreement between
a UK operator and a UK University group. The agreement defines what
data may be collected; what uses it may be put to; how privacy of the
data originators is to be protected and that any machines storing
data must be protected to the standard of best practice for their
operating environments.  The detailed text of these issues has been
based on a framework document previously generated for this purpose by
JANET(UK)\footnote{\url{http://www.ja.net/documents/development/legal-and-regulatory/regulated-activities/traffic-data-for-research.doc}}.
The legal aspects of this agreement were made considerably simpler by
the aims of the MASTS project to record only protocol information from
the Transport Layer and below. As such, no Application Layer data are
stored and hence no data produced directly by a user (such as email text)
is collected. Privacy is however still potentially compromised by the
presence of the Network Layer (\ie IP) Address. The agreement therefore
requires such addresses to be anonymised in such a way that end user
privacy is maintained.  Technical solutions to this issue are discussed
in Section~\ref{sec:anon}.

\item\label{en:agg2} A legal agreement which allows MASTS users at
other institutions to be registered as visitors to the UCL network.
This allows the cover provided by agreement~\ref{en:agg1} to extend to
them if they are named in this agreement.

\item\label{en:agg3} A legal agreement between the site holding the
data and non-MASTS users.  This is similar to the other agreements and
ensures that data are not disclosed to third parties or used for purposes
other than those agreed. In addition, users must agree to acknowledge
the source of the data in any work published and not attempt to reverse
the anonymisation.
\end{enumerate}

The project has also established a dissemination approach which does
not require data users to sign a legal agreement.  This approach makes
available summary data in which individual packet-level data is not
available. For example, total data rate, or the number of different
source IP addresses (but not the anonymised IP addresses themselves)
within a given period of time can be provided world wide via the web
interface without the establishment of a formal agreement. The only
requirement made of a user is that of acknowledging the source of the
data and using this for approved purposes only.

\subsection{Data Available}
\label{sec:dataavailable}

\begin{figure}[ht!]
\begin{center}
\includegraphics[width=0.9\textwidth]{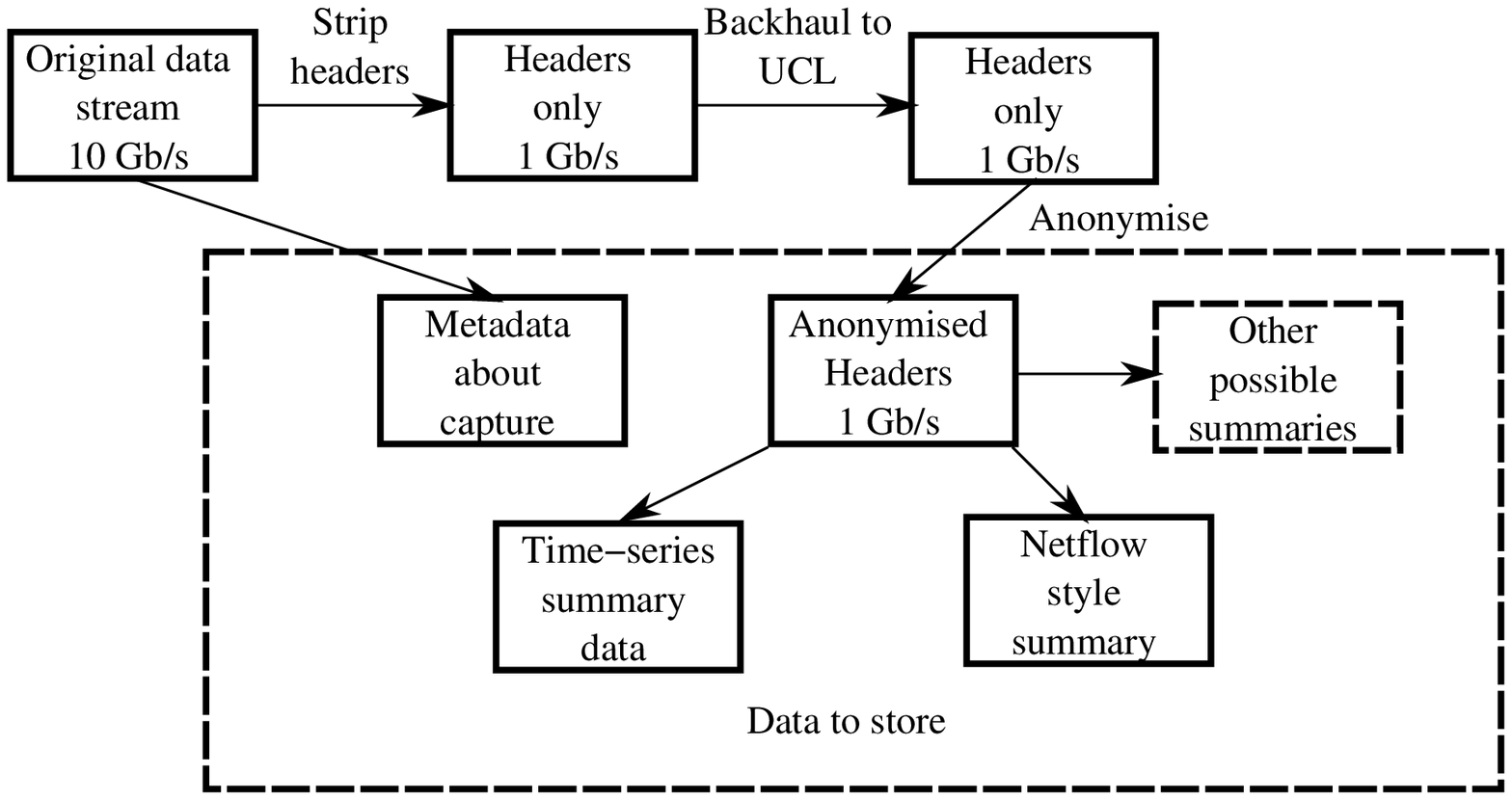}
\caption{The path of the data from monitoring point to database.}
\label{fig:datapath}
\end{center}
\end{figure}

Figure~\ref{fig:datapath} shows the path of the data and the various
transformations which occur between the monitoring point and the database.
The initial traffic streams are expected to have a maximum rate of
10Gb/s.  The traffic is split as described in Section~\ref{sec:physarc}
and only the headers retained.  The data is anonymised as described
in Section~\ref{sec:anon}.  Metadata about the capture process and
extracted summary data is placed in a searchable database as described
in Section~\ref{sec:dataarc}.  The data is provided in the Extensible
Record Format (ERF) which has high timestamp fidelity and includes
loss information~\cite{donnelly02high}.  Tools are also be provided to
convert the data to the more common \emph{pcap} format commonly used by
researchers\footnote{\url{http://research.wand.net.nz/software/libtrace.php}}.

Obviously with such a high data arrival rate the data store would fill quickly.
Tests have been performed on several large trace files to estimate this.
The data sets considered here include a 500GB data set covering 24 hours 
from the site-connection of a medium size research institute and some
typical data sets (each approximately 10GB) collected in 2002 and downloaded
from the CAIDA website.  In the first data set, stripping headers
reduced the data to 14\% of its original volume.  Compression techniques
on the headers (gzip and lzo were both tried in --best and --fast mode) 
reduced the headers further to between 4.5\% and 6.1\% of the 
original volume depending on the technique used.  
Using standard parameters, taking netflow 
style summary data without sampling reduced the data to  1.2\% of the original
volume and taking 1/512 packet samples reduced that data to 0.0071\% of the
original volume.

Table \ref{tab:datatypes} shows how quickly various summary methods would
fill 10TB of storage which represents the amount of storage that this
project could reasonable devote to storing a single type of data
from one monitoring point.  The table shows the full data, the headers only,
the headers compressed using gzip (the differences between the various
compression algorithms tried were quite small), Netflow data without
sampling and netflow data using 1/512 packet sampling.  The figures are
based on the assumption that on average the data arrives in the system
at 10\% of the maximum system capacity (that is, the data is arriving at
a mean rate of 1Gb/s rather than the maximum rate 10Gb/s).  The figures
are given to only a single figure of accuracy and are based upon the
results of the previous paragraph.  It is obvious that for all but the
most extremely compressed data storage formats storage can only be for
a limited time period.  Those extremely compressed formats, however,
carry much less information.  For example, one of the options is to
store the number of bytes of data seen in every millisecond interval as a
time-series for the lifetime of the project.  However, the research value
of this data is much less than the research value of full header data.

The final solution which is used for the MASTS project is to have several
levels of data kept.  Extremely summarised data (for example bytes seen
in a given time unit) can be stored for the lifetime of the project.
Complete header information is stored for a short period for those
researchers who wish to look at the current day of traces or who might
want to examine traces to investigate a particular special event which
has recently occurred on the network.  A small repository of complete
header files is kept for a longer time period.  This repository
will be useful for researchers who want representative traces to test
data analysis schemes or hypotheses about, for example, creation of
synthetic traffic traces.  Finally, representative metadata (such as
sampled netflow) may be stored for a longer time period, which will be
determined by the amount of storage space taken up by that data.

\begin{table}
\begin{center}
\begin{tabular}{|c | c c c |} \hline
Data  & Max  & Mean  & Time to  \\
Format & rate & rate & fill 10TB \\ \hline
Full data & 10Gb/s & 1Gb/s &  1 day \\
Headers &  1Gb/s & 100Mb/s &  1 week \\
Comp. headers &  500Mb/s& 50Mb/s &  2 weeks \\
Full netflow &  100Mb/s & 10Mb/s  & 3 months \\
1/512 netflow &  700Kb/s &70Kb/s & 30 years \\ \hline
\end{tabular}
\caption{Types of data which might be stored with approximate data rates and
estimated time to fill 10TB of storage.}
\label{tab:datatypes}
\end{center}
\end{table}

\section{Conclusions and Future Work}
\label{sec:conc} 

While the MASTS project does not finish until 2009, considerable
progress has been made.  Obviously, the laws applying to such data
collection vary considerably across jurisdictions, the legal framework
given here would be directly useful to those considering monitoring
in the UK and could be a model to adapt for those in other countries.
We consider this legal framework an important outcome of the project
which could be useful to other monitoring researchers.

The difficulty of a monitoring project of this type should not be
underestimated.  We hope that the experiences described in this paper will
provide a useful guide for those considering attempting such a project.
As described in Section~\ref{sec:operational}, there are several
concerns which may cause problems and delay in monitoring projects.
In specific, monitoring equipment must be deployed with minimum harm
to network availability, equipment may use protocols which differ from
those established and delays in scheduling the installation of equipment
can cause difficulties.

Data sets are to be made available from the project website:
\url{http://www.mastsproject.org/}.  These data sets will be valuable
to networking research.  The ability to monitor recent traces from the
JANET network will allow researchers to save data sets of particular
value and when network events of interest occur.  The utility of a
monitoring project is best judged by the research it stimulates and it
is hoped that the data provided here will be of considerable use both
in the understanding it will bring and in the new research opportunities
it will provide.

The MASTS project has provided a combination of tools both legal and
engineering, as well as encouraging the operational relationships to ease
future monitoring, particularly at the large scale. It is clear that the
monitoring systems in place within MASTS may be easily extended to cover
larger aspects both of the JANET interconnect to the internet and across
the regions of the JANET infrastructure. There is no reason to be limited
to the JANET networks and with the great interconnection diversity in
the UK, provided by many broadband providers and peering locations such
as LINX\footnote{https://www.linx.net/} (the London Internet Exchange),
this will lead to a rich and diverse set of monitoring opportunities.

As an extension to the basic search functions and visualisation
of the metadata, more flexible preprocessing and advanced
visualisation~\cite{withall2007} of the data will be developed.
The extensions to the data processing will partly be based on the idea of
storing intermediate information~\cite{bashir1998} and also incorporate
ideas from other network data processing work \eg\cite{fisk2002}.
In addition, caching of downloaded trace files may be incorporated as
part of the web server to minimise read load on the archive.

\subsection*{Thanks}

The JANET Lightpath network was funded by JISC and HEFCE.  MASTS was
funded under the EPSRC grants GR/T10503/01, GR/T10510/03, and
GR/T10527/01.

We would also like to thank Andrew Cormack and Richard Clayton who
contributed invaluable expertise in the preparation of the legal documents
and David Miller for his contributions of technical expertise.


\bibliographystyle{abbrv}
\bibliography{topology,db,legal}
\end{document}